\pdfoutput=1
\documentclass[twocolumn,superscriptaddress,aps,preprintnumbers,
prl,nofootinbib
]{revtex4-2}

\usepackage{graphicx}
\usepackage{xcolor}
\usepackage[caption=false]{subfig}
\usepackage{mathrsfs,mathtools}
\usepackage{physics,amssymb}
\usepackage{siunitx}
\usepackage{bm}
\usepackage{braket}
\usepackage{listings}
\usepackage{cases}
\usepackage{comment}
\usepackage{soul}
\usepackage{cancel}
\usepackage{cases}
\usepackage[utf8]{inputenc}
\usepackage{url}
\usepackage{longtable}
\usepackage{xspace}
\usepackage{acronym}
\usepackage{aas_macros}

\usepackage{amsfonts}
\usepackage{xfrac}
\usepackage{pifont}
\usepackage{fourier}
\usepackage{hyperref}
\usepackage{bm}
\usepackage{enumitem}

\definecolor{rossoferrari}{HTML}{D9073D}
\definecolor{mediumblue}{HTML}{0000CD}
\definecolor{forestgreen}{HTML}{228B22}
\definecolor{desy_blue}{HTML}{009EE2}
\definecolor{desy_orange}{HTML}{FD8800}
\definecolor{light_pink}{rgb}{1,0.4,0.4}
\definecolor{light_blue}{rgb}{0.284602,0.317763,0.963947}
\hypersetup{setpagesize=false,bookmarksnumbered=true,bookmarksopen=true,%
colorlinks=true,linkcolor=light_blue,urlcolor=rossoferrari,citecolor=rossoferrari,linktocpage=false}

\newcommand{\ee}{\mathrm{e}}
\newcommand{\Mpl}{M_\mathrm{Pl}}
\newcommand{\ns}{n_{\mathrm{s}}}

\newcommand{\uc}{\mathrm{c}}

\newcommand{\Ne}{
N}
\newcommand{\D}{\mathscr{N}}
\newcommand{\HI}{H_{\rm inf}}

\newcommand{\bae}[1]{\begin{align} #1 \end{align}}

\definecolor{MONZA}{HTML}{CF000F}
\definecolor{DARKBLUE}{HTML}{00008b}
\definecolor{DARKMAGENTA}{HTML}{8b008b}

\def\beqa{\begin{eqnarray}}
\def\eeqa{\end{eqnarray}}

\newcommand{\GeV}{\ {\rm GeV}}

\def\lmk{\left(}
\def\rmk{\right)}

\def\la{\left<}
\def\ra{\right>}
\newcommand{\eq}[1]{Eq.~(\ref{#1})}
\def\beq#1\eeq{\begin{align}#1\end{align}}

\newcommand{\curv}{\mathcal{R}}
\newcommand{\Nc}{\Ne_{\rm c}}

\begin{document}

\preprint{TU-1230}

\title{
Multifield Stochastic Dynamics in GUT Hybrid Inflation 
\\
and Gravitational Wave Signatures of GUT Higgs Representation
}

\author{Yuichiro Tada}
\email{tada.yuichiro.y8@f.mail.nagoya-u.ac.jp}
\affiliation{Institute for Advanced Research, Nagoya University,
Furo-cho Chikusa-ku, 
Nagoya 464-8601, Japan}
\affiliation{Department of Physics, Nagoya University, 
Furo-cho Chikusa-ku,
Nagoya 464-8602, Japan}

\author{Masaki~Yamada}
\email{m.yamada@tohoku.ac.jp}
\affiliation{Department of Physics, Tohoku University, Sendai, Miyagi 980-8578, Japan}
\affiliation{FRIS, Tohoku University, Sendai, Miyagi 980-8578, Japan}

\date{\today}

\begin{abstract}
\noindent
We revisit the hybrid inflation model within the framework of the Grand Unified Theory (GUT), focusing on cases where the waterfall phase transition extends over several e-foldings to dilute monopoles. Considering the stochastic effects of quantum fluctuations, we demonstrate that the waterfall fields (i.e., GUT Higgs) maintain a nonzero vacuum expectation value around the waterfall phase transition. By accurately accounting for the number of degrees of freedom of the GUT Higgs field, we establish that these fluctuations can produce observable gravitational waves without leading to an overproduction of primordial black holes. The amplitude of these gravitational waves is inversely proportional to the degrees of freedom of the waterfall fields, thereby providing a unique method to probe the representation of the GUT Higgs.
\end{abstract}\acresetall

\maketitle


\acrodef{CMB}{cosmic microwave background}

\smallskip\noindent{\textbf{Introduction.\,---\,}} 
The Grand Unified Theory (GUT) is a compelling candidate for extending the Standard Model (SM) of particle physics~\cite{Georgi:1974sy,Fritzsch:1974nn,Pati:1974yy,Mohapatra:1974hk,Mohapatra:1974gc,Senjanovic:1975rk,Gursey:1975ki}. It notably offers a straightforward framework for hybrid inflation associated with GUT phase transitions. It is important to note that 't Hooft--Polyakov monopoles~\cite{tHooft:1974kcl,Polyakov:1974ek} arise during spontaneous symmetry breaking (SSB) per unit correlation volume, potentially leading to a monopole overproduction problem~\cite{Guth:1981uk}.

To circumvent the issue of monopole overproduction, one might consider a symmetry-breaking pattern that does not result in monopole generation~\cite{Watari:2004xh}, such as flipped GUT~\cite{Kyae:2005nv,Rehman:2009yj,Hertzberg:2014sza,Rehman:2018nsn}. This can occur through the SSB of SO(10), assumed to take place prior to inflation, thereby sufficiently diluting any monopoles~\cite{Dvali:1994ms,Ellis:2014xda}. Another approach involves monopole production at an intermediate scale, as seen with SU(4)$\times$SU(2)$\times$SU(2)~\cite{Senoguz:2015lba} or the elimination of monopoles by domain walls~\cite{Dvali:1997sa,Bachmaier:2023zmq}. In Ref.~\cite{Shafi:1983bd}, a coupling between the inflaton and GUT Higgs was explored, stabilizing the GUT Higgs at the SSB minimum. 
Smooth hybrid inflation, where the GUT symmetry breaks spontaneously before inflation, has also been considered~\cite{Lazarides:1995vr,Jeannerot:2000sv,Antusch:2010va,Khalil:2010cp}. A related, yet distinct mechanism is subcritical hybrid inflation, where inflation extends for more than 60 e-folds post-waterfall phase transitions to dilute monopoles~\cite{Buchmuller:2014dda,Bryant:2016tzg}. 
Additionally, the potential role of the GUT Higgs as the inflaton, which would involve breaking the GUT symmetry during inflation, has been discussed~\cite{Senoguz:2004ky,Ijaz:2023cvc}.

This study explores scenarios with substantial e-folding following during the waterfall phase transition, under the assumption of a relatively flat potential for the inflaton near the critical point. Given that the GUT symmetry is spontaneously broken and monopoles are produced around this phase transition, their abundance can be significantly reduced. However, careful attention is needed for the non-trivial dynamics of the GUT Higgs field during the extended waterfall phase due to large quantum fluctuations induced by stochastic effects~(see Refs.~\cite{Starobinsky:1982ee,Starobinsky:1986fx,Nambu:1987ef,Nambu:1988je,Kandrup:1988sc,Nakao:1988yi,Nambu:1989uf,Mollerach:1990zf,Linde:1993xx,Starobinsky:1994bd} for the first papers on the stochastic effect in inflation).

Inflaton potentials similar to those discussed here have been considered in different contexts in Refs.~\cite{Clesse:2010iz,Kodama:2011vs,Clesse:2015wea}. The role of quantum fluctuations in numerous waterfall fields has been analyzed, with varying degrees of accuracy in handling stochastic effects~\cite{Halpern:2014mca,Tada:2023fvd}. We extend the concept to multiple waterfall fields charged under gauge symmetry. 
Such stochastic behavior is critical around the potential's origin, namely the symmetry-enhancement point, where the Higgs potential is primarily quadratic. At this juncture, all degrees of freedom contribute to the dynamics, except for the modes absorbed into the massive gauge field. The stochastic dynamics are particularly pronounced for a large number of waterfall fields, that is, a large representation of GUT Higgs~\cite{Tada:2023fvd}.

The phase transition during inflation generate large curvature perturbations on a small scale, resulting in observable induced gravitational waves (GWs)~\cite{Saito:2008jc, Saito:2009jt}. Our model and calculations, which are applicable to all GUT groups, predict varying amplitudes for these induced GWs, thus providing a novel method to probe the representation of the GUT Higgs field.

{\it Note added: 
As we were finalizing our paper, a related study~\cite{Moursy:2024hll} was published on arXiv. It discusses a similar approach to GUT hybrid inflation to address the monopole overproduction issue, focusing on a specific GUT model but without considering stochastic effects.
}



\smallskip\noindent{\textbf{GUT hybrid inflation.\,---\,}} 
We consider the hybrid inflation model, where inflation ends with the waterfall phase transition at a critical point $\phi_\uc$ for the inflaton field $\phi$. 
The waterfall phase transition is identified as the spontaneous symmetry breaking of the GUT symmetry, and the waterfall fields are identified as the GUT Higgs $\Psi$. 
The analysis in this study can be applied to any GUT model, but depends on the total number of real degrees of freedom for the GUT Higgs, $\D$. 
Let us first review the representation of the GUT Higgs in representative GUT models to determine a typical value of $\D$.

In the simplest SU(5) model, the SU(5) gauge symmetry is spontaneously broken into the SM gauge group $G_{\rm SM}$ by $\bm{24}_H$. 
One may extend the model to address the doublet-triplet splitting problem using for example, the missing partner mechanism, in which case the GUT symmetry may be spontaneously broken by $\bm{ 75}_H$~\cite{Masiero:1982fe, Grinstein:1982um, Yamada:1992kv, Hisano:1994fn, Altarelli:2000fu}. 
In the SO(10) GUT model, one can consider the symmetry-breaking pattern of, for example, SO(10) (or spin(10)) $\to$ SU(4)$_c \times$SU(2)$_L \times$SU(2)$_R$ using $\bm{54}_H$ or $\bm{210}_H$ at the GUT scale~\cite{Lazarides:1982tw,Holman:1982tb}. 
Two GUT Higgs can be introduced to complement Higgs with real representations~\cite{Reiss:1981nd}. In addition, in supersymmetric models, one may have to introduce two Higgs to cancel the gauge anomaly. 
In general, higher or many representations for the GUT Higgs can be introduced; however, one should consider a Landau pole to ensure perturbativity above the GUT scale. Therefore, we expect the number of degrees of freedom for the GUT Higgs to be $\D \sim 100$.

For illustrative purposes, let us briefly consider the Higgs field $\Psi$ in the adjoint representation under SU($5$), such that 
\beq
 V(\Psi) = \Lambda^4 -\frac12 m^2 \Tr \Psi^2 + \frac14 \lambda_1 \Tr \Psi^4 + \frac14 \lambda_2 \lmk \Tr \Psi^2 \rmk^2 + \frac13 A \Tr \Psi^3 \,.
 \label{eq:Vpsi}
\eeq
During the waterfall phase transition, the self-interaction terms can be neglected because $\la \Psi \ra \ll v_{\rm GUT}$, where $v_{\rm GUT}$ ($\sim 10^{16}\GeV$) represents the GUT scale. 
Consequently, the relevant term in the scalar potential is reduced to a quadratic term, which is universal for any gauge symmetry. Therefore, we can generically consider the following potential: 
\beq
 V(\Psi) \simeq \Lambda^4 -\frac12 m^2 \sum_i \psi_i^2 \,, 
\eeq
where $\psi_i$ ($i = 1,2,\dots, \D$) represents each component of the GUT Higgs field. 
The following discussion is based only on the quadratic term of the Higgs potential, which generally exhibits approximate O($\D$) symmetry in the absence of gauge symmetry. Therefore, our scenario can be applied to GUT Higgs with any representation for any GUT model.

As an order of magnitude estimation, the GUT scale is given by $v_{\rm GUT}^2 = \la \sum \psi_i^2 \ra \sim m^2 / \lambda$, where $\lambda$ represents a typical value of $\lambda_i$. 
The potential energy at the origin of the potential $V_0$ is determined such that the potential energy at the potential minimum vanishes. This implies $\Lambda \sim m / \lambda^{1/4} \sim \lambda^{1/4} v_{\rm GUT}$.

To realize the waterfall phase transition, we introduce a coupling between the inflaton $\phi$ and GUT Higgs $\Psi$, e.g., $\sum_i \psi_i^2 \phi^2$, such that 
the GUT Higgs remains at the origin of the potential for $\phi > \phi_\uc$ 
and has a tachionic mass for $\phi < \phi_\uc$. 
We are interested in the dynamics around the waterfall phase transition $\phi = \phi_\uc$. The potentials for the inflaton and waterfall fields can be represented as 
\bae{
    V(\phi,\bm{\psi})=\Lambda^4\bqty{
    1
    +2 \lmk \frac{ \phi^2 - \phi_\uc^2 }{\phi_\uc^2}  \rmk \frac{\sum_i \psi_i^2}{M^2} +\frac{\phi-\phi_\uc}{\mu_1}-\frac{(\phi-\phi_\uc)^2}{\mu_2^2}
    + \dots
    } \,,
    \label{eq:potential}
}
where we define $M^2 \equiv 4 \Lambda^4/m^2$ ($\sim v_{\rm GUT}^2$). 
The dots represent higher-order terms for the inflaton potential as well as other potential terms for the waterfall fields that are negligible for the dynamics around the waterfall phase transition. 
We denote $\sum_i \psi_i^2 = \psi_r^2$ by defining the radial direction in the field space of waterfall fields as $\psi_r$.

\smallskip\noindent{\textbf{Inflaton dynamcis.\,---\,}} 
First, we consider the dynamics along the inflaton direction $\phi$, which can be analyzed without considering the stochastic dynamics. 
The potential energy around the critical point is almost constant, and is approximated as follows: 
$V \simeq \Lambda^4$. 
The Universe then experiences inflation with the Hubble parameter $H \simeq \Lambda^2/(\sqrt{3} \Mpl)$, where $\Mpl$ is the reduced Planck mass. 
The potential curvature along the waterfall fields vanishes at $\phi = \phi_\uc$, which we call the critical point for the waterfall phase transition and denote its e-folding number from the beginning of inflation as $\Nc$. 
The time evolution of the inflaton is given by 
$\phi \simeq \phi_\uc - \Mpl^2 (\Ne-\Nc)/ \mu_1$ for $\Ne \approx \Nc$, where $N$ denotes the forward e-folding number as a time variable. 

For a sufficiently large $\mu_1$, 
inflation can last a few e-foldings even after $\phi$ reaches the critical point. 
We define the end of inflation as 
$\abs{V_{\psi \psi} \Mpl^2/V } = 1$
or 
$(\phi_\uc - \phi_{\rm end})/\phi_\uc \simeq M^2/(8 \Mpl^2)$ ($\ll 1$). 
The e-folding number between the critical point for the waterfall phase transition and the end of inflation $\Delta N_{\rm PT}$ 
is calculated as 
\bae{
    \Delta N_{\rm PT} \sim \frac{\Pi^2}{8} \,,
} 
where we define~\cite{Clesse:2015wea}
\bae{
    \Pi &\equiv \frac{M \sqrt{\mu_1 \phi_\uc}}{\Mpl^2} 
    \simeq 8 \lmk \frac{M}{10^{16} \GeV} \rmk 
    \lmk \frac{\mu_1}{4 \times 10^{6} \Mpl} \rmk^{1/2}
    \lmk \frac{\phi_\uc}{\Mpl} \rmk^{1/2} \,,.
    \label{eq:Pi}    
}
As we will see shortly, we are interested in the case where $\Pi = \mathcal{O}{(10})$ to dilute monopoles enough.

The observed amplitude of the curvature perturbations at the pivot scale $k_*$ ($= 0.05 \, {\rm Mpc}^{-1}$) is given by 
${\cal P}_\curv (k_*) \simeq 2.1 \times 10^{-9}$~\cite{Planck:2018vyg}. 
In our case, the corresponding scale exits the horizon before the waterfall phase transition, and the amplitude is calculated from ${\cal P}_\curv = \HI^2 / (8 \pi^2 \epsilon \Mpl^2)$ with $\epsilon 
 \simeq \Mpl^2/(2 \mu_1^2)$. 
We therefore require 
$\Lambda \simeq 2.7 \times 10^{13} \GeV 
 / \sqrt{\mu_1/(4 \times 10^6 \Mpl)}$. 
The spectral index is calculated as $\ns \simeq 1 - 4  \Mpl^2/\mu_2^2$.
We assume $\mu_2 \simeq 10 \Mpl$ to explain the observed spectral index of $\ns = 0.9649 \pm 0.0042$~\cite{Planck:2018vyg}.


\smallskip\noindent{\textbf{Stochastic dynamics for waterfall fields.\,---\,}} 
When the parameter $\Pi^2$ is as large as or larger than $\mathcal{O}(10)$, 
the stochastic effects of quantum fluctuations during inflation becomes important for waterfall fields. 
Owing to the Goldstone Boson Equivalence Theorem, the longitudinal mode of gauge bosons in the Higgs phase is equivalent to a massless scalar in the high-energy limit. 
Therefore we can consider the stochastic effect in these fields to be similar to the case of global symmetry, at least when the gauge boson mass is considerably lower than the Hubble parameter during inflation~\cite{Graham:2015rva,Sato:2022jya}. 
We can thus use the formalism of the stochastic effects for multiple light fields explained in Ref.~\cite{Tada:2023fvd} in the regime of $g \psi_r \ll H$, where $g$ ($< 1$) is the gauge coupling constant at the GUT scale.

In the regime where the mass of the waterfall field is small and $g \psi_r \ll H$, 
the Langevin equation for the radial direction $\psi_r$ is 
\bae{
        \partial_\Ne \psi_r=-\frac{4 \psi_r \Mpl^2 }{M^2} \lmk \frac{\phi^2 - \phi_\uc^2}{\phi_\uc^2} \rmk +\frac{1}{2} \lmk \frac{H}{2\pi}\rmk^2 \frac{\D-1}{\psi_r}+\frac{H}{2\pi}\xi_{\psi_r}(\Ne) \,,
    \label{eq:stochastic_r}
}
where the second term on the right-hand side represents the effective centrifugal force in the radial direction in the field space $\psi_i$ due to the stochastic noise. 
The noise term obeys 
$\braket{\xi_{\psi_r}(\Ne)\xi_{\psi_r}(\Ne^\prime)}=\delta(\Ne-\Ne^\prime)$. 
Using $\phi \simeq \phi_\uc - \Mpl^2 (\Ne-\Nc)/\mu_1$, 
we obtain 
the evolution equation for $\braket{\psi_r^2}$ such that
\bae{
    \dv{\braket{\psi_r^2}}{\Ne} = \pqty{\frac{4}{\Pi}}^2 (\Ne-\Nc) \braket{\psi_r^2} + \D \frac{V}{12\pi^2\Mpl^2} \,. 
}
We should note that this expression for $\braket{\psi_r^2}$ holds only when the Higgs is light enough, i.e., $\abs{V_{\psi\psi}\Mpl^2/V}\lesssim1$. 
We therefore set 
an initial condition of 
$\braket{\psi_r^2} = 0$ at $\Ne - \Nc = -\Delta N_{\rm PT}$ and use the above equation for $\Ne - \Nc \ge -\Delta N_{\rm PT}$.

Once $g \psi_r$ becomes significantly larger than $H$, the gauge fields gain mass and can be integrated out for our purposes. Subsequently, the effective degrees of freedom for the Higgs field, $\D$, should be replaced by $\D'$ in the Langevin equation, 
where $\D'$ ($\le \D$) denotes the number of degrees of freedom that takes into account the reduction due to the absorption into the massive gauge boson. 
Although the details in the intermediate regime remain unclear, we anticipate that these two limiting regimes are smoothly connected around $g \psi_r \sim H$. Specifically, we propose that the curvature perturbations for the waterfall fields are calculated using multifield stochastic equations with $\D'$, particularly for the e-foldings around $\Ne \sim \Nc$.

Here we note the e-folding number at which the GUT symmetry is spontaneously broken. We expect that the GUT symmetry will not be restored at a later epoch once the typical amplitude of each component, $\abs{\psi_i}$, exceeds $H/2\pi$. 
According to the Langevin equation, this occurs when
$(\Ne - \Nc) = \mathcal{O}(1)$. (Although the amplitude of the radial direction becomes as large as $\sqrt{\D H/2\pi}$, the typical amplitude of each component of the Higgs field, $\psi_i$, is of the same order as its fluctuations for every Hubble time.) Therefore, we conclude that the GUT symmetry is spontaneously broken at the e-folding number $\Ne \sim \Nc$.

\smallskip\noindent{\textbf{Monopole abundance.\,---\,}} 
Inflation stretches the correlation length for the field value of the GUT Higgs. 
The correlation length of the GUT Higgs is given by $\zeta H_I^{-1}$, 
where $\zeta \simeq \ee^{\Delta N_{\rm SSB}}$ 
at the end of the inflation period and we manifest the Hubble scale $H_I$ during inflation by the subscript $I$. 
After the GUT symmetry breaking, 
a monopole is expected to be produced per unit correlation volume, 
and its number density at the end of inflation is estimated to be $n_M \sim 4 \pi \zeta^{-3} H_I^3/3$. 
The monopole number density over the current entropy density is then given by 
\beq
 \frac{n_M}{s} 
 &\simeq \frac{\pi \zeta^{-3} T_{\rm RH} H_I}{3 \Mpl^2} \,. 
 \label{eq:YM}
\eeq
Here, $T_{\rm RH}$ is the reheating temperature after inflation, which depends on the GUT model but has an upper bound such that $T_{\rm RH}  \lesssim  8.4 \times 10^{13} \GeV  \sqrt{ H_I/10^{10} \GeV}$,
where we use $g_* \simeq 106.75$ for the effective relativistic degrees of freedom at the end of reheating. 
In the simplest GUT model, 
the GUT Higgs can quickly decay into the light Higgs, 
and the reheating temperature is expected to be comparable to the maximum value.

Several experiments and observations place the upper bound on the monopole abundance or flux, including 
the Parker bound~\cite{Parker:1970xv,Turner:1982ag,Adams:1993fj,Lewis:1999zm}, 
the monopole track search in ancient mica~\cite{Price:1983ax},
the MACRO experiment~\cite{MACRO:2002jdv},
nucleon-decay search experiments~\cite{Kajita:1985aig,Bartelt:1986cv,Becker-Szendy:1994kqw,Baikal:1997kuo,MACRO:2002iaq,IceCube:2014xnp},
the upper bound of the neutrino flux from the sun~\cite{Super-Kamiokande:2012tld}, 
and other astrophysical constraints~\cite{Kolb:1982si,Harvey:1982py,Dimopoulos:1982cz,Arafune:1983tr,Freese:1983hz,Kolb:1984yw,Freese:1998es} (see Ref.~\cite{ParticleDataGroup:2022pth} for details). 
If one adopts the constraint from Ref.~\cite{Super-Kamiokande:2012tld}, 
it requires $n_M/s \lesssim 10^{-39}$ or $\zeta \gtrsim 10^{9} \simeq \ee^{21}$ for $H_{\rm inf} = 10^{10} \GeV$ from \eq{eq:YM}. 
This implies that the GUT symmetry should be spontaneously broken about $\Delta N_{\rm mono} \simeq 21$ e-foldings before the end of inflation. 
If the extended Parker bound is instead adopted, it can be addressed by $\Delta N_{\rm mono} \simeq 15$.
To address the monopole overproduction problem, we require 
$\Delta \Ne_{\rm PT} \gtrsim \Delta N_{\rm mono}$ or $\Pi \gtrsim 11 ( \Delta N_{\rm mono}/15)^{1/2}$.

\smallskip\noindent{\textbf{Induced gravitataional waves.\,---\,}} 
The stochastic fluctuations of waterfall fields result in large curvature perturbations on a small scale. This phenomenon has been analyzed both analytically and numerically in Ref.~\cite{Tada:2023fvd}, including the dependence on the number of waterfall fields. The peak amplitude and the number of e-foldings are given by
\beq
 &\mathcal{P}_{\mathcal{R}}^{\rm (peak)} \simeq 
 0.0057 \lmk \frac{\Ne_{\rm PT}}{15} \rmk 
 \lmk \frac{\D'}{10} \rmk^{-1} \,, 
 \label{eq:peakP}
 \\
 &\Delta \Ne_{\rm PT} \sim 15 \lmk \frac{\Pi^2}{115} \rmk^{1/2}\,.
\eeq
Note that the latter quantity is consistent with the above analytic estimation within a $10\%$ precision.

Importantly, the peak amplitude is reduced by a factor of $1/\D'$ for a large representation of the GUT Higgs field. 
This occurs because the Langevin equation, \eq{eq:stochastic_r}, introduces a type of centrifugal force in the field space when $\D' \ge 2$, resulting in relatively smaller perturbations relative to the overall field value amplitude. This suppression factor is crucial for phenomenology. Without this factor, the peak amplitude could reach as high as $0.04$, which would predict an excessive formation of primordial black holes (PBHs) from collapsing overdense regions in a radiation-dominated epoch. 
It should also be noted that the abundance of PBHs can be as large as that of dark matter in cases where $\D' \sim 5$ and $\Pi^2 \sim 100$, with the corresponding PBH mass being around $10^{20} \, {\rm g}$.

\begin{figure}
    \centering
            \includegraphics[width=0.95\hsize]{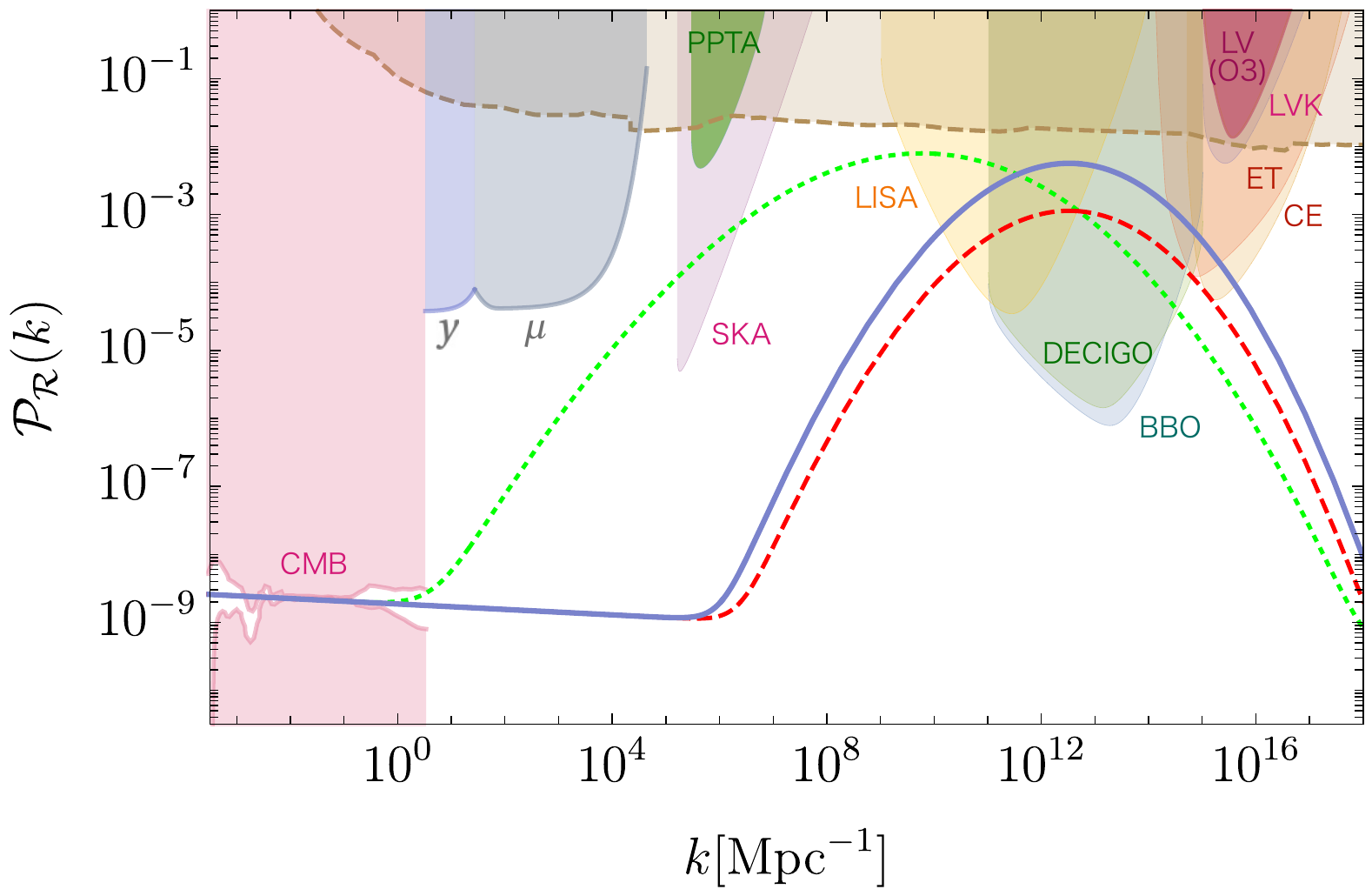}
    \caption{
    Spectrum of curvature perturbations generated by the GUT hybrid inflation model for cases: $(\D', \Pi^2) =(10,115)$ (blue solid curve), ($10, 225$) (green dotted curve), and ($50, 115$) (red dashed curve). 
}
    \label{fig:1}
\end{figure}

The detailed spectrum for curvature perturbations is depicted in Fig.~\ref{fig:1}:
the blue curve represents the case of $\D' =10$ and $\Pi^2 = 115$, 
and the green dashed curve represents $\Pi^2 = 225$.
We also plot the case of $\D' = 50$ and $\Pi^2 = 115$ as the red dashed curve.
Constraints from the CMB observations are shown in the lightly shaded region for $k \lesssim 10^4 \, {\rm Mpc}^{-1}$~\cite{Nicholson:2009pi, Nicholson:2009zj, Bird:2010mp, Bringmann:2011ut,Fixsen:1996nj,Chluba:2012we,Bianchini:2022dqh}. 
Constraints from the overproduction of PBHs are indicated by the dashed curve in the upper region~\cite{Raidal:2017mfl,Serpico:2020ehh,Inoue:2017csr,Page:1976wx,Carr:2009jm,Boudaud:2018hqb,Laha:2019ssq, DeRocco:2019fjq,Zumalacarregui:2017qqd,Niikura:2017zjd,Blaineau:2022nhy,Wyrzykowski:2011tr,Niikura:2019kqi} (see Ref.~\cite{Escriva:2022duf} for details).

Large scalar perturbations induce GWs through second-order effects~\cite{Saito:2008jc, Saito:2009jt} (see also Refs.~\cite{Espinosa:2018eve,Kohri:2018awv} for useful analytic formulas and Ref.~\cite{Domenech:2021ztg} for a recent review). 
Current constraints from PTA experiments~\cite{Shannon:2015ect} and direct GW detection experiments~\cite{KAGRA:2021kbb} are shown in the figure as densely shaded regions. Future sensitivity curves for upcoming GW detection experiments are presented in lighter shaded areas, including 
SKA~\cite{Janssen:2014dka},
LISA~\cite{LISA:2017pwj},
DECIGO~\cite{Kawamura:2011zz,Kawamura:2020pcg},
BBO~\cite{Harry:2006fi},
Einstein Telescope (ET)~\cite{Punturo:2010zz,Maggiore:2019uih},
Cosmic Explorer (CE)~\cite{Reitze:2019iox},
and aLIGO+aVirgo+KAGRA (LVK)~\cite{Somiya:2011np,KAGRA:2020cvd}.
We anticipate that our predictions will be testable by these GW experiments. In particular, 
the value of $\D'$ can be determined from observations of the GW spectrum, as its peak frequency depends on $\Delta \Ne_{\rm PT}$ and the peak amplitude on both $\Delta \Ne_{\rm PT}$ and $\D'$. This offers a novel method to verify the representation of the GUT Higgs field through GW detection experiments.

\smallskip\noindent{\textbf{Discussion and conclusions.\,---\,}} 
We have revisited a GUT hybrid inflation model, considering the stochastic effects on the GUT Higgs field. 
During inflation, the field value of the GUT Higgs tends to be larger for a larger representation of the GUT Higgs due to quantum fluctuations. 
The stochastic effect leads to large curvature perturbations on a small scale. 
Employing a stochastic equation for a single waterfall field would result in curvature perturbations so large that they lead to an overproduction of PBHs. Based on our recent study, we have correctly accounted for the effect of a large number of degrees of freedom for the GUT Higgs field, showing that the PBH abundance is significantly reduced and consistent with observations.

Furthermore, we have demonstrated that induced GWs from large scalar perturbations can be within the observable range for future GW detection experiments. The amplitude of these GWs is inversely proportional to the degrees of freedom of the GUT Higgs, providing a unique method to verify the representation of the GUT Higgs field. Moreover, since the relevant number of degrees of freedom $\D'$ is almost uniquely determined for each GUT model, we can also determine the gauge group of the GUT model.

Motivated by recent observation of stochastic GWs by pulsar-timing array (PTA) experiments~\cite{NANOGrav:2023gor,Reardon:2023gzh,EPTA:2023fyk,Xu:2023wog}, 
scenarios involving 
metastable cosmic strings have been extensively considered in the literature~\cite{Leblond:2009fq,Buchmuller:2019gfy,Buchmuller:2020lbh,Buchmuller:2021mbb,Lazarides:2023rqf,Buchmuller:2023aus,Antusch:2023zjk,Ahmed:2023rky,Lazarides:2023ksx,Fu:2023mdu,Maji:2023fhv,Afzal:2023cyp,Afzal:2023kqs,Ahmed:2023pjl,Servant:2023tua,Chitose:2023dam,Lazarides:2024niy}.
(See also Refs.~\cite{Yamada:2022aax,Yamada:2022imq,Yamada:2023thl} for cosmic superstring interpretation of PTA data.) 
Our scenario of GUT hybrid inflation naturally realizes cosmic string formation post-inflation without a monopole overproduction problem. 

Other interesting signals are imprinted in the cosmic microwave background (CMB) temperature anisotropies via a mechanism known as the cosmological collider~\cite{Chen:2009zp,Baumann:2011nk,Noumi:2012vr,Arkani-Hamed:2015bza}. 
The waterfall fields possess a time-dependent effective mass around the pivot scale, which provides a characteristic signature in the bispectrum~\cite{Wang:2018tbf,Aoki:2023wdc}.

Finally, we comment on the effect of the cubic term on the waterfall field and the domain wall problem. 
If we do not impose $Z_2$ symmetry, then 
a cubic term for the GUT Higgs (similar to the last term in \eq{eq:Vpsi}) may be introduced. 
Its effect on the curvature of the potential is of the order of $\Lambda^4 \psi_r / M^3$. 
Since this is much smaller than $H^2$ during the waterfall phase transition, 
our calculations were not affected by this cubic term for the parameter of interest. 
Later, the cubic term became effective and biased the vacuum energy to address the domain-wall problem. 
To address this issue, we did not need to fine-tune the cubic term.


\medskip\noindent\textit{Acknowledgments\,---\,}%
YT is supported by JSPS KAKENHI Grants 
No.~JP21K13918 and No.~JP24K07047.
MY was supported by MEXT Leading Initiative for Excellent Young Researchers, and JSPS KAKENHI. Grant Nos.\ JP20H05851 and JP23K13092.


\bibliographystyle{JHEP}
\bibliography{ref}


\end{document}